\documentclass[a4paper,12pt]{article}
\usepackage{amsmath,amssymb,amsfonts,epsfig,graphicx}
\usepackage{colordvi}
	\fboxrule 0.2mm 	\newcounter{comentario} 	

\title{Berry's phase in  the two-level model}

\author{
 A.C. Aguiar Pinto$^{1}$, M. Moutinho$^{1}$ and 
 M.T. Thomaz$^{2}\footnote{Corresponding author: mtt@if.uff.br}$   \vspace{0.25cm} \\ 
\small\it $^{1}$
Curso de F\'{\i}sica, Universidade Estadual de Mato Grosso do Sul,\\
\small\it Caixa Postal: 351, Cidade Universit\'aria de Dourados, CEP 79804-970, 
Dourados-MS,  Brazil.
\vspace{0.25cm}\\
\small\it $^{2}$Instituto de F\'{\i}sica, Universidade Federal Fluminense,\\ 
\small\it Av. Gal. Milton Tavares de Souza s/n$^{\textit o}$, 
CEP 24210-346, Niter\'oi-RJ, Brazil
}

\begin{document}

\maketitle

\begin{abstract}

We study the adiabatic evolution of a two-level model
in the presence of an external classical electric field. The
coupling between the quantum model and the classical field 
is taken in the electric dipole approximation.  In this regime,
we show the absence of geometric phases in the interacting 
two-level model in the presence of any periodic real 
time-dependent classical electric field. We obtain a conservative
scalar potential  in the calculation of Berry's phases
of the instantaneous eigenstates of the model. For complex electric
fields, we recover the existence of geometric phases. In particular,
the geometric phases of the instantaneous eigenstates of the model 
in the presence of a positive or of a negative frequency component 
of the  monochromatic electric field differ by an overall sign.
As a check on our results, we map this interacting 
two-level model onto a spin-$1/2$ model under
the action of a classical magnetic field.  We confirm that the 
first one acquires  Berry's phase only in the rotating wave 
approximation [RWA].

\end{abstract}

\vfill



\newpage


\section{Introduction} \label{S1}

A wide application of quantum systems with two energy levels can be 
found in Physics. One important example
is the spin-$1/2$ model coupled to an external magnetic 
field, which describes the spin-$1/2$ magnetic dipole in 
nuclear magnetic  resonance\cite{rabi1936,rabi1937}. Another 
example  is the quasi-resonant regime of an atom under the
action of a monochromatic laser whose wavelenght is much larger than
the average electronic radius. The latter phenomenon is described by
the rotating wave approximation (RWA) of the two-level
model\cite{livro_optica}. In this approximation, the anti-resonant part
of the probability transition is neglected, which corresponds to
replacing the monochromatic electric field by its positive 
frequency component.

The two-level models have the nice feature of being exactly soluble
and hence good toy models. They can be used to understand the
conditions under which a given approximation can be applied. In
Berry's original paper about the geometric phases \cite{berry}, he
showed how those phases are acquired by instantaneous eigenstates of
the spin-1/2 model adiabatically evolving in the presence of an
external periodic classical magnetic field. In Ref.\cite{ajp2000} we
studied the spin-$1/2$ hamiltonian, and derived the exact
instantaneous eigenstates of this model in the presence of a
precessing classical magnetic field with fixed norm and compared them
to the approximate eigenstates obtained from the adiabatic
theorem\cite{fock,marcelo}.

Since the publication of Ref.\cite{berry}, 
discussion on the presence of geometric phases in the
two-level system derived from the coupling of matter with a
monochromatic electric  field \cite{livro_optica} can 
be found in the literature. Its RWA was discussed
in Refs. \cite{nao-hermitiana,surya}; an irreversible version 
is presented in Ref. \cite{andrei}; and a periodically 
driven two-model with time-dependent diagonal elements 
can be found in \cite{kayanuma}. There has been a recent
revival of the two-level  model, as it has been 
suggested as a prototype for a quantum bit (qubit).
In 2003, Leibfired  {\it et al}. used geometric phases 
associated to the  two-model model, in a quasi-resonant 
regime, of  an atom under the action of a 
monochromatic laser  to  model a logic  
gate\cite{leibfried}. The quantum state has to have a
cyclic evolution\cite{aharonov} in the presence of an
periodic monochromatic external electric field, that has not
adiabatic evolution, since it has to be in the quasi-resonant
regime  for the two-level model to describe to interaction
of matter and radiation.
	
The aim of the present communication is to study the 
adiabatic evolution of the instantaneous eigenstates 
of a two-level quantum hamiltonian with a adiabatically-varying
periodic external classical electric field, in the electric 
dipole approximation.  In section
\ref{S2} we calculate explicitly the geometric phases of 
this model from the expression given by 
Berry\cite{berry}, and explore some consequences 
for real and complex effective classical electric 
fields.  In section \ref{S3}, for the sake of verification, 
we map this model onto the spin-$1/2$ model under
the action of an external classical magnetic field to 
discuss the presence/absence of Berry's phase in the 
instantaneous eigenstates of the two-level model for a 
time-dependent classical electric field, including the 
RWA of the model. Finally, in section \ref{S4} we 
summarize our results for two-level hamiltonian with
an arbitrary periodic time-dependent external electric 
field.
 

\section{The interacting two-level model} \label{S2}

The hamiltonian ${\bf H}_0$ of the quantum two-level model,
without interaction, written in terms of its eigenstates is

\begin{eqnarray} \label {2.1}
{\bf H}_0 = \varepsilon_1 |1\rangle \langle 1| 
           + \varepsilon_2 |2\rangle \langle 2| , 
\end{eqnarray}

\noindent where ${\bf H}_0 |i \rangle = \varepsilon_i |i \rangle$,
$i= 1,2$. We consider $\varepsilon_1 < \varepsilon_2$, which
prevents (\ref{2.1}) from describing a degenerate
two-level system.

A first motivation to study a two-level model comes from
an original  quantum system confined to a spatial region of
finite diameter, interacting with a classical time-dependent 
external electric field $\vec{E} (t)$, the wavelength of 
which being much larger than the diameter of the region and 
in the quasi-resonant regime. This last two conditions 
allow us to use the electric dipole approximation for the 
interaction in the RWA.  The interaction hamiltonian 
${\bf H}_e (t)$,  represented in the basis of the eigenstates 
of ${\bf H}_0$, is \cite{lamb}

\begin{eqnarray}  \label{2.2}
  {\bf H}_e (t) = \left(
     \begin{array}{c c}
     \varepsilon_1  &  (\vec{d}_{12} \cdot \vec{E} (t))^{*} \\
      \vec{d}_{12} \cdot \vec{E} (t) & \varepsilon_2
     \end{array}
  \right) ,
\end{eqnarray}

\noindent  in which

\begin{eqnarray}  \label{2.3}
  \vec{d}_{12} \equiv -e  \, \langle 1| \vec{\bf x} | 2 \rangle ,
\end{eqnarray}

\noindent and $\vec{\bf x}$ is the position operator of the electron.
We assume that the eletronic states $|i \rangle$, $i= 1,2$,
are spherically symmetric, and hence 
do not have a permanent electric dipole ($\vec{d}_{ii} =0$, $i=1,2$).
We are using  the  natural units, $\hbar= c=1$.

The components of the vector $\vec{d}_{12}$ can be 
imaginary numbers, depending on the choice of the 
basis $\{ |i \rangle, i=1,2 \}$ and they are
time-independent.  Throughout our calculations, 
we have chosen the basis of eigenstates of ${\bf H}_0$, 
such that  the components of  $\vec{d}_{12}$ are 
real; we have taken $e>0$, so that the electronic 
charge is $-e$. 

The RWA of hamiltonian (\ref{2.2}) has been used in 
quantum computation to implement geometric phase
gates between ion-qubits in the presence of an external
electric field, whose wavelength are much larger than
the space distribution of the atom and it is in the
quasi-resonant regime\cite{leibfried,tian,imai1, imai2}.
The geometric operations associated to the time evolution
of hamiltonian (\ref{2.2}) are considered to be potentially
robust\cite{tian}.

\vspace{0.3cm}

For the sake of future comparison between the adiabatic 
evolution of  the two-level hamiltonian (\ref{2.2}) and 
that of another two-level model that describes the 
interaction of the spin-1/2 with an  external magnetic 
field\cite{berry}, we make a trivial shift in the 
hamiltonian and define

\begin{eqnarray}  \label{2.4}
{\bf H}_e^{\prime} (t) \equiv  {\bf H}_e (t)
     - \frac{(\varepsilon_1 + \varepsilon_2)} {2} {\bf 1},
\end{eqnarray}

\noindent  in which ${\bf 1}$ is the  $2\times 2$ 
identity matrix. In order to simplify the mathematical 
expressions  in ${\bf H}_e^{\prime} (t)$ we define 

\begin{subequations}

\begin{eqnarray}
	\Delta \varepsilon \equiv \varepsilon_2 - \varepsilon_1,   \label{2.5a} \\
%
%
 D(t) \equiv \vec{d}_{12} \cdot \vec{E} (t) .  \label{2.5b}
\end{eqnarray}

\end{subequations}

\noindent We finally have

\begin{eqnarray}  \label{2.6}
  {\bf H}_e^{\prime} (t) = \left(
     \begin{array}{c c}
     - \frac{ \Delta \varepsilon}{2} &  (D (t))^{*} \\
         &     \\
      D (t) & \frac{ \Delta \varepsilon}{2}
     \end{array}
  \right) .
\end{eqnarray}

From now on, we study the time evolution of the quantum system
driven by a general hamiltonian of the type  (\ref{2.6}) when it evolves
adiabatically. We have $\Delta \varepsilon \in \mathbb{R}$ and 
\mbox{$D(t) \in \mathbb{C}$;} the latter function 
has a period $T$ satisfying  the adiabatic  condition 
\mbox{$T \cdot \left(\frac{\Delta \varepsilon}{ 2\pi}\right) \gg1$.}
Any function $D(t)$ can be written  in the form (\ref{2.5b}); 
the effective external electric field $\vec{E} (t)$ 
evolves adiabatically and is out of the quasi-resonant 
regime. In order to have a slow-varying periodic 
function $D(t)$, it is sufficient that the effective 
external field have an adiabatically-evolving
periodic component along the direction of the 
time-independent  vector $\vec{d}_{12}$. We
continue to assume that the hamiltonian (\ref{2.6}) is 
non-degenerate at any instant $t$. 

Following Berry's original notation \cite{berry}, 
in the general case we have a 
hamiltonian that varies in time due to 
a time-dependent classical field $\vec{R} (t)$, that is,
${\bf H} (t) = {\bf H} (\vec{R} (t))$. The classical
field $\vec{R} (t)$ is supposed to evolve adiabatically.
The instantaneous eigenvalue equation  of energy is

\begin{eqnarray}   \label{2.7}
{\bf H} (\vec{R} (t)) | n; \vec{R} (t) \rangle  
    = \lambda_n (\vec{R} (t)) | n; \vec{R} (t) \rangle ,
\end{eqnarray}

\noindent where $| n; \vec{R} (t) \rangle$ is the 
instantaneous eigenvector of hamiltonian 
${\bf H} (\vec{R} (t))$ with eigenvalue
$\lambda_n (\vec{R} (t))$. The instantaneous
spectrum $\{ \lambda_n (\vec{R} (t))\}$ is supposed
to be non-degenerate. From Ref.\cite{berry}, each
eigenstate $| n; \vec{R} (t) \rangle$ 
acquires the following  Berry's phase:

\begin{eqnarray}  \label{2.8}
\gamma_n = -i \oint_{\Gamma_{\vec{R}}} \, 
  \langle n; \vec{R} (t)| \vec{\nabla}_{\vec{R}} (| n; \vec{R} (t) \rangle)
     \cdot d\vec{R}, 
\end{eqnarray}
                
\noindent  in which $\Gamma_{\vec{R}}$ is an oriented 
closed loop in the  $\vec{R}$-parameter space.

In the two-level hamiltonian (\ref{2.6}), coupled to an 
effective  classical  electric field, 
[cf. eqs.(\ref{2.2})-(\ref{2.6})] we have 
$\vec{R} (t) = \vec{E} (t)$. In order  to include the RWA 
of this hamiltonian, we consider the general case
where the components of the effective classical field, 
vector $\vec{E} (t)$, can  be complex.

From now on, we have

\begin{eqnarray}  \label{2.9}
{\bf H} (\vec{R} (t)) = {\bf H}_e^{\prime} (\vec{E} (t)) .
\end{eqnarray}

The time dependence  of the hamiltonian ${\bf H}_e^{\prime}$,
of the function $D$ and of the eigenvalues $\lambda_n$ 
is solely due to the  time dependence of the 
classical electric field. In order to 
simplify the notation, we write this operator
and functions as  ${\bf H}_e^{\prime} (t)$,  $D (t)$
and  $\lambda_n (t)$, respectively.

The hamiltonian ${\bf H}_e^{\prime} (t)$ has two instantaneous
eigenvalues:

\begin{eqnarray}  \label{2.10}
 \lambda_{l} (t) = l \;  \sqrt{\left(\frac{\Delta \varepsilon}{2}\right)^2 
        + |D(t)|^2} \, , 
\hspace{0.5cm}  l= \pm 1 .
\end{eqnarray}

The instantaneous normalized eigenstates of ${\bf H}_e^{\prime} (t)$,
associated to $\lambda_{l} (t)$, for $l = \pm 1$, are

\begin{subequations}

\begin{eqnarray}
W_{l} (t) & = & \left( 
         \begin{array}{c}
         (A_{l} (t))^{1/2} \\
                   \\

       \frac{l}{\sqrt{2}} {D(t)}{ (F_{l} (t))^{-1/2}}
         \end{array} \right)  \label{2.11a} \\
 &&  \nonumber  \\
 &&  \nonumber  \\
%
%
  & \equiv & \left(
     \begin{array} {c}
      u_{l} (t) \\
             \\
      v_{l} (t)
     \end{array} \right)  ,   \label{2.11b}
\end{eqnarray}

\end{subequations}

\noindent so that

\begin{eqnarray} \label{2.12}
A_{l} (t) \equiv \frac{1}{2} - l \, \cdot \frac{\Delta \varepsilon}{4} \,
     \frac{1} {\sqrt{ \left(\frac{\Delta \varepsilon}{2}\right)^2 + |D(t)|^2}}
\end{eqnarray}

\noindent and 

\begin{eqnarray}  \label{2.13}
F_{l} (t) = \left(\frac{\Delta \varepsilon}{2}\right)^2  
+ |D(t)|^2  - l \, \cdot  \left(\frac{\Delta \varepsilon}{2}\right) \,
\sqrt{ \left(\frac{\Delta \varepsilon}{2}\right)^2 + |D(t)|^2} .
\end{eqnarray}

Using the entries  of $W_{l} (t)$, after a closed
loop in the $\vec{E}$-parameter space, we can write
Berry's phase $\gamma_{l}$ (see eq. (\ref{2.8})) as

\begin{eqnarray}  \label{2.14}
\gamma_{l} = -i \oint_{\Gamma_{\vec{E}}} \,
\left[ u_{l}^{*} (\vec{\nabla}_{\vec{E}} u_{l})  
+ v_{l}^{*} (\vec{\nabla}_{\vec{E}} v_{l})   \right] \cdot d\vec{E} , 
\mbox{\hspace{0.5cm} for \hspace{0.5cm}} l= \pm 1 .
\end{eqnarray}

We now calculate $\gamma_{+1}$ explicitly. Since

\begin{eqnarray}  \label{2.15}
\frac{\partial u_{+1}}{\partial E_i} = \frac{1}{2} \,
\frac{1}{(A_{+1})^{1/2}} \,  \frac{\partial A_{+1}}{\partial E_i}, 
     \hspace{1cm} 
	i \in \{ x, y, z\},
\end{eqnarray}

\noindent we obtain 

\begin{eqnarray}  \label{2.16}
u_{+1}^{*} (\vec{\nabla}_{\vec{E}} u_{+1}) 
= \vec{\nabla}_{\vec{E}} \left[\frac{A_{+1} (t)}{2}\right] ,
\end{eqnarray}

\noindent that is a gradient in $\vec{E}$-parameter 
space.

In the general case (including the RWA of hamiltonian
${\bf H}_e^{\prime} (t)$) we have that $D(t) \in \mathbb{C}$,
then 

\begin{eqnarray}  \label{2.17}
  D(t) = |D(t)| e^{i \phi (t)} ,
\end{eqnarray}

\noindent where $\phi (t)$ is a real phase. Substituting 
eq.(\ref{2.17})  in $v_{+1} (t)$  yields

\begin{eqnarray} \label{2.18}
v_{+1} (t) = \frac{|D (t)|}{\sqrt{2}} \, \frac{e^{i \phi(t)}}{(F_{+1} (t))^{1/2}} .
\end{eqnarray}

The derivative of $v_{+1} (t)$ with respect to the components of the 
external electric field $\vec E$ is

\begin{eqnarray} \label{2.19}
\frac{\partial v_{+1}}{\partial E_i} = \frac{e^{i \phi(t)}}{\sqrt{2}}
\left[
\frac{1}{F_{+1}^{1/2}} \,\frac{\partial [|D(t)|]}{\partial E_i}
- \frac{|D(t)|}{2 F_{+1}^{3/2}} \,\frac{\partial [ F_{+1} (t)]}{\partial E_i}
+ \frac{i \, |D(t)|}{F_{+1}^{1/2}} \,\frac{\partial [\phi(t)]}{\partial E_i}
\right]  ,
\end{eqnarray}

\noindent in which $i \in \{ x, y, z\}$. Remembering that 
$v_{+1} (t) \in \mathbb{C}$, we have

\begin{eqnarray} \label{2.20}
v_{+1}^{*} \, (\vec{\nabla}_{\vec{E}} v_{+1})  = 
\vec{\nabla}_{\vec{E}} \left[\frac{|D(T)|^2}{4 F_{+1}(t)} \right]
+ \frac{i}{2} \frac{|D(T)|^2}{F_{+1}(t)} \left[ \vec{\nabla}_{\vec{E}}(\phi(t))\right].
\end{eqnarray}

Substituting results (\ref{2.16}) and (\ref{2.20}) on the r.h.s.
of eq.(\ref{2.14})  for $l= +1$, it becomes

\begin{eqnarray} \label{2.21}
\gamma_{+1} = -i \oint_{\Gamma_{\vec{E}}} \,
\left[ 
\vec{\nabla}_{\vec{E}} \left[\frac{A_{+1} (t)}{2} \,
+ \, \frac{|D(T)|^2}{4 F_{+1}(t)} \right]
+ \frac{i}{2} \frac{|D(T)|^2}{F_{+1}(t)} \left[ \vec{\nabla}_{\vec{E}}(\phi(t))\right]
\right] \cdot d\vec{E} .
\end{eqnarray}

The integral of the first term on the r.h.s. of expression 
(\ref{2.21})  vanishes for a closed loop $\Gamma_{\vec{E}}$. 
Rewriting the second term on the r.h.s. of this expression, 
we obtain

\begin{eqnarray}  \label{2.22}
\gamma_{+1} = \frac{1}{2} 
\int_{0}^{T} dt \, \; \frac{|D(t)|^2} {F_{+1}(t)} \; \dot{\phi} (t)  ,
\end{eqnarray}

\noindent where $T$ is the period of the function 
$D(t)$ ($D(0) = D(T)$) and 
$\dot{\phi} (t) \equiv \frac{d\phi(t)}{dt}$.

The calculation of $\gamma_{-1}$ is analagous and we 
obtain, for $l = \pm 1$,

\begin{eqnarray} \label{2.23}
\gamma_{l} = \frac{l}{2} 
\int_{0}^{T} dt \, \; \frac{|D(t)|^2} {F_{l}(t)} \; \dot{\phi} (t) . 
\end{eqnarray}

\vspace{0.3cm}

A few comments on the geometric phase of the instantaneous
eigenstates of ${\bf H}_e^{\prime} (t)$ (see hamiltonian (\ref{2.6}), 
or (\ref{2.2}) when it is written in terms of an effective 
classical  electric field, are in order.

\begin{itemize}
\item [\it i)] From the result (\ref{2.16}), we verify that 
the entry $u_l (t)$ in the eigencolumns of $W_l (t)$, 
for $l = \pm 1$, does not contribute to the geometric 
phase $\gamma_l$.

\item[\it ii)] For $D(t) \in \mathbb{R}$ (or, equivalently, 
$\vec{E} (t) \in \mathbb{R}^3$) the phase $\phi (t)$ is 
time-independent, and ${\bf H}_e^{\prime} (t)$ is a 
one-parameter hamiltonian. Therefore its instantaneous 
eigenstates  do not acquire a geometric phase along a closed 
loop in the $\vec{E} (t)$-space. This last statement agrees 
with the result derived from eq. (\ref{2.23}) for real $D(t)$.

\item[\it iii)] For $D(t) \in \mathbb{C}$ we can have a 
non-null geometric phase $\gamma_l$, $l= \pm 1$, since 
${\bf H}_e^{\prime} (t)$ becomes a two-parameter 
hamiltonian. This case corresponds to the RWA of 
${\bf H}_e^{\prime} (t)$ when we have 
$\vec{E} (t) \in \mathbb{C}^3$.

\item[\it iv)] Two complex fields $\vec{E}_1 (t)$ 
and $\vec{E}_2 (t)$ with the same component along the 
vector $\vec{d}_{12} (t)$  have the same geometric phase 
$\gamma_l$, $l = \pm 1$.

\item[\it v)] For complex fields $\vec{E} (t)$ it suffices 
to have a periodic component of this field along 
$\vec{d}_{12} (t)$ for the instantaneous eigenstates of 
${\bf H}_e^{\prime} (t)$ to acquire a geometric phase.

\item[\it vi)] For a complex monochromatic classical electric 
field,

	\begin{eqnarray}  \label{2.24}
	 \vec{E}_{\pm} (t) = \vec{E}_0 \, e^{\mp i (\omega t + \phi_0)}  ,
	\end{eqnarray}
 
\noindent in which $\omega$ is the angular frequency of the 
classical field, $\vec{E}_0$ is its real amplitude  and 
$\phi_0$ is a real constant phase. From expression (\ref{2.23}) 
we obtain the relation between the geometric phases 
$\gamma_l^{(+)}$ and $\gamma_l^{(-)}$, respectively related to the 
monochromatic electric fields $\vec{E}_{+} (t)$ and $\vec{E}_{-} (t)$, 
which reads

	\begin{eqnarray}   \label{2.25}
	\gamma_l^{(+)} = - \gamma_l^{(-)}, \hspace{1cm} {\rm for} \;\; l= \pm1.
	\end{eqnarray}

Relation (\ref{2.25}) comes from the linear dependence on $D(t)$ 
in the numerator of the entry $v_l (t)$ (see eq. (\ref{2.11a})). 
As a consequence of this result we obtain that for 
a hamiltonian (\ref{2.2}) in the presence of a classical
field of the form 
\mbox{$\vec{E} (t) = \vec{E}_{+} (t) + \vec{E}_{-} (t)$,}
its instantaneous eigenstates do not acquire a 
geometric phase.

\item [\it vii)] Finally, we verify the absence of Berry's 
phase in a two-level quantum system coupled to any real 
classical external electric field, in the electric dipole 
approximation and in the adiabatic regime. This result comes 
from the fact that for this  hamiltonian (\ref{2.2}) we have
$\langle n; \vec{R} (t)|
\vec{\nabla}_{\vec{R}}| (|n; \vec{R} (t)\rangle) =
\vec{\nabla}_{\vec{E}} [ G(\vec{E})]$, that corresponds to a scalar
potencial $G(\vec{E})$ of a conservative system for any time-dependent
real classical electric field, thus yielding null integrals for closed
paths in $\vec{E}$-parameter space.

\end{itemize}


\vspace{0.5cm}

\section{Comparison of two-level models} \label{S3}

In his seminal paper\cite{berry}, Berry showed that geometric phases
are acquired by the instantaneous eigenstates of a spin-$1/2$ particle
under a precessing magnetic field $\vec{B}(t)$ in adiabatic evolution,
described by the hamiltonian

\begin{eqnarray} \label{3.1} 
{\bf H}_{spin} (t) = \mu \vec{\sigma} \cdot \vec{B}(t), 
\end{eqnarray} 

\noindent in which
$\mu= \frac{1}{2} g \mu_B$, $g$ is the Land\'e 
factor and $\mu_B$ is the Bohr magneton. 

In order to compare the adiabatic evolution of the two-level 
models ${\bf H}_e^{\prime} (t)$ (see eq. (\ref{2.4})) and 
${\bf H}_{spin} (t)$ (eq. (\ref{3.1})) we use the Pauli 
matrices to rewrite hamiltonian (\ref{2.4}) as

\begin{eqnarray}  \label{3.2}
{\bf H}_e^{\prime} (t) =  \Re [ \vec{d}_{12} \cdot \vec{E} (t)] \sigma_x 
  + \Im [ \vec{d}_{12} \cdot \vec{E} (t)] \sigma_y
  + \frac{(\varepsilon_1 - \varepsilon_2)} {2} \sigma_z,
\end{eqnarray}

\noindent  in which $\Re[...]$ and $\Im[...]$  stand for 
the real and imaginary parts of the expression in brackets,
respectively. 

Comparing hamiltonians (\ref{3.1}) and (\ref{3.2}), we obtain 
that ${\bf H}_e^{\prime} (t)$ is equivalent to the spin
hamiltonian in the presence of an external classical magnetic
field with components

\begin{subequations}

\begin{eqnarray} 
 B_x(t) & \equiv & \frac{1}{\mu} \Re[ \vec{d}_{12} \cdot \vec{E} (t)], 
                                   \label{3.3a}  \\
%
%
B_y(t) & \equiv & \frac{1}{\mu} \Im[ \vec{d}_{12} \cdot \vec{E} (t)], 
                                   \label{3.3b}  \\
%
%
B_z(t) & \equiv & \frac{\varepsilon_1- \varepsilon_2}{2 \mu}.   \label{3.3c}
\end{eqnarray}

\end{subequations}

In the general case,  we use the definition (\ref{2.5b}) to rewrite the 
scalar product 

\begin{subequations}

\begin{eqnarray} 
 \vec{d}_{12} \cdot \vec{E} (t) &\equiv & D(t)  \\
%
%
           & \equiv & |D(t)| e^{i \phi (t)} .  \label{3.4}
\end{eqnarray}

\end{subequations}

Substituting expression (\ref{3.4}) in the components 
of the equivalent magnetic field, we obtain

\begin{subequations}

\begin{eqnarray}
 B_x (t) &=& \frac{|D(t)| }{ \mu} \cos [\phi(t)] ,  \label{3.5a} \\
%
%
B_y (t) &=& \frac{|D(t)| }{ \mu} \sin [\phi(t)] ,  \label{3.5b} \\
%
%
B_z(t) & = & \frac{\varepsilon_1- \varepsilon_2}{2 \mu}  .  \label{3.5c}
\end{eqnarray}

\end{subequations}

These equations describe an external magnetic field with a constant
$z$-component. For any real vector $\vec{E} (t)$ we have 
$B_y (t) = 0$ and the equivalent magnetic field 
varies along a straight line in the $(x,z)$ plane. The path
being a straight line in the parameter space of the magnetic field
yields a null solid angle, and hence produces no geometric phase to
the instantaneous eigenstates of ${\bf H}_e(t)$\cite{berry}. Such
absence of a geometric phase happens for a quantum system under a
classical external eletric field $\vec{E}$ that follows a closed loop
in its parameter space. This result is valid for any periodic
classical electric field with real components. 

Although the hamiltonian of a two-level model 
must be hermitian, we have an important application 
of this model in  quantum optics, namely for a 
monochromatic time-dependent external electric 
field $\vec{E} (t)$ in the quasi-resonant 
regime\cite{livro_optica}. Under these conditions 
the dynamics of the electronic part
of the quantum model is well approximated by the 
RWA of the hamiltonian  (\ref{2.2}). In this 
approximation, the main contribution
to the transition probability amplitude  comes
from the positive frequency of $\vec{E}_+ (t)$ 
\mbox{(see eq. (\ref{2.24})).} in this case, he 
components of the effective magnetic field a time 
dependent phase $\phi (t)$ given by

\begin{eqnarray}  \label{3.6}
 \phi(t) \equiv -(\omega t +  \phi_0) . 
\end{eqnarray}

The previous magnetic field configuration describes a precessing
field with a non-null solid angle. The instantaneous eigenstates 
of the RWA of the two-level model acquire a geometric phase 
under an adiabatic  periodic evolution of the 
monochromatic
electric field \nolinebreak(\ref{2.24}), 
which is already known in the  literature\cite{nao-hermitiana}.

The results of section \ref{S3} fully agree with the consequences
of expression (\ref{2.23}) for  Berry's phases $\gamma_l$,
$ l= \pm 1$, associated to the instantaneous eigenstates of 
hamiltonian ${\bf H}_e^{\prime}$.


\section{Conclusions} \label{S4}

We study the adiabatic evolution of a two-level quantum model
driven by hamiltonians of type (\ref{2.6}), which 
appears in the interaction of matter with 
a classical electric field in the quasi-resonant 
regime\cite{livro_optica}. The interaction
is taken in the electric dipole approximation and
the hamiltonian is handled in RWA. The RWA 
of hamiltonian (\ref{2.2}) has been used 
in the modelling of prototype systems for qubits 
in quantum  computation, constituted by atoms 
interacting with electric fields\cite{leibfried}.

By explicitly calculating  
Berry's phases of 
the instantaneous eigenstates of hamiltonian 
${\bf H}_e^{\prime} (t)$, see eq. (\ref{2.4}), we 
show that they are absent when the quantum system 
is under the action of any time-dependent  real 
classic  electric field that evolves adiabatically. 
This result is a direct consequence of the 
quantum system being driven by a one-parameter 
hamiltonian.  For any real time-dependent electric 
field,  the geometric phase is calculated over
a closed loop of a gauge field with a conservative 
potential,  thus giving a null result.
Certainly, this result is unexpected since it is valid
for any real time-dependent  electric field, 
which in general  is a non-conservative field.

On the other hand, we can have geometric phases acquired
by the instantaneous eigenstates of the RWA of 
hamiltonian  ${\bf H}_e^{\prime} (t)$; a sufficient 
condition is that the component of the classical 
electric field along the direction of $\vec{d}_{12}$, 
defined by  expression (\ref{2.3}), be periodic and 
evolve adiabatically. 

A very interesting result comes from the $\phi$-dependence
of  Berry's phase (eq. (\ref{2.23})). Interactions with 
the positive-frequency part $\vec{E}_+ (t)$ and
negative-frequency part $\vec{E}_- (t)$ of a real monochromatic
electric field yield contributions to the geometric phase that differ
only by a global sign. 
If the quantum system interacts with the (real) superposition 
of these two components of $\vec E$ the geometric phases of the 
instantaneous eigenstates of the interaction hamiltonian vanish. 
This set of consequences derived from eq. (\ref{2.23}) rise the 
question  whether the two-level model (\ref{2.2}) is suitable 
for the calculation of geometric phases in the interaction 
of atoms  with  monochromatic laser fields, in the 
quasi-resonant regime\cite{leibfried,tian,imai1,imai2}
or whether the original hamiltonian for the atom with an 
infinite number of bound states in the presence of a 
real monochromatic electric field should be used instead. 
In order to verify the correction of the results of 
section \ref{S2},  hamiltonian (\ref{2.2})  is  mapped onto a 
spin-$1/2$  model coupled to an external magnetic 
field\cite{nao-hermitiana}. The presence of geometric phases in 
this last model was showed by Berry in his seminal paper\cite{berry}. 
In Ref.\cite{nao-hermitiana}  Garrison and Wright showed that
the instantaneous eigenstates of the RWA of the two-level
model in the presence of a monochromatic  classical electric 
field acquire geometric phases.

\vspace{1cm}

The authors are in debt with Sir Michael V. Berry for his comments 
on the original version of the present communication. He raised
the question of the validity of using the RWA of hamiltonian
(\ref{2.2}) to calculate the geometric phases of atoms in the 
presence of  a monochromaticelectric field in the quasi-resonant 
regime. 

M.T. Thomaz  (Fellowship CNPq, Brazil, Proc.No.: 30.0549/83-FA) 
thanks CNPq  for partial financial support.


\end{document}